\documentstyle[12pt,preprint,aps]{revtex}
\tighten
\draft
\begin{document}
\title{Integral equation theory for the electrode-electrolyte interface
       with the central force water model. Results for an aqueous solution
       of sodium chloride }

\author{M. Vossen, F. Forstmann \\
        Institut f{\"u}r Theoretische Physik, Freie Universit{\"a}t Berlin \\
        Arnimallee 14, 14195 Berlin, Germany}
\date{\today}
\address{Submitted to The Journal of Chemical Physics }
\maketitle
\renewcommand{\baselinestretch}{1.4}

\begin{abstract}
The structure of an aqueous solution of sodium chloride at a
planar electrode is investigated by integral equation techniques. With the
central force water model the aqueous electrolyte is modelled as a mixture
of sodium and chloride ions and partially charged hydrogen and oxygen atoms
interacting via effective spherically symmetric pair potentials. The
correlation functions obtained from the Ornstein-Zernike equation with
Reference-Hypernetted-Chain closure gives a good description of the bulk
structure (e.g. hydrogen bonded water network, solvation shell). With the
bulk information and the Wertheim-Lovett-Mou-Buff equation we have calculated
the density profiles at the uncharged and charged electrode. The rather rigid
ice-like water structure found previously at the neutral surface strongly
repels the ions. Steric interactions between the differently sized ions and
the ice-like water structure dominates the ionic distribution near the
electrode. This model electrolyte also responds differently to opposite
charges on the electrode. We found the asymmetry in the differential
capacitance curve entirely determined by the response of the interfacial
water structure.
\end{abstract}
\pacs{}
\renewcommand{\baselinestretch}{1.5}

\section{introduction}
\noindent
In a previous article \cite{vos94} we have described an integral equation
method for calculating accurate bulk correlation functions for the central
force (CF) water model and reported results about the water structure at an
electrode. The central force water model \cite{lem75}, which represents water
as a stoichiometric mixture of partially charged hydrogen and oxygen atoms
interacting via spherically symmetric pair potentials, allows a
completely molecular description of water within the framework of integral
equation theory. The effective pair potentials include the information about
the proper triangular arrangement of atoms within the water molecule, the
molecular dipole moment and the formation of hydrogen bonds. This water model
has been successfully employed and optimized in computer simulations
\cite{sti78,bop83}. The water structure obtained near the electrode
has been interpreted in terms of an ice-like arrangement of water molecules in
good agreement with several computer simulation \cite{lee84} and integral
equation results \cite{tor88}. In this paper we extend our previous
investigations to an aqueous model electrolyte and consider the consequences
of the interfacial water structure on the structure of the electrical double
layer \cite{vos94a}.

Early theoretical investigations of the electrical double layer neglected
the molecular nature of the solvent. The solvent was approximated as a
dielectric continuum and solvent effects appear only in mediating the
Coulombic interactions. A mean-field treatement of this primitive model
electrolyte leads to the Gouy-Chapman theory for the double layer, which
holds in the dilute regime where the diffuse part of the electrical double
layer is dominated by the long ranged Coulomb interactions. The remaining
inner layer adjacent to the electrode was assumed to be formed by solvent
molecules and excluding nonspecifically adsorbed ions. Several
phenomenological models for the water monolayer employing adjustable
parameters for the specific water interaction with the electrode were
proposed to account for the experimental data \cite{gui86,boc93}. Of
particular interest is the pronounced asymmetry of the differential
capacitance with respect to the potential of zero charge, which characterizes
the interface between an aqueous electrolyte and a mercury electrode
\cite{gra52}.

A step further towards a microscopic description of the
liquid side of the interface was achieved by employing integral equation
methods based on the statistical mechanics of liquids \cite{blu92} or on
density functional arguments \cite{eva92,alt87}. First of all, in the
'primitive model' of an electrolyte the ions were treated as hard spheres with
a point charge in the center. Then the continuum solvent
was replaced by molecular models. Even for the simplest model electrolyte of
equal sized hard sphere ions and dipolar hard spheres a detailed fluid
structure emerges throughout the interface as a consequence of both solvent
and ionic interactions \cite{ber91,don89,dia94}. With respect to the
tetrahedral charge distribution within a water molecule, Torrie et al.
supplemented the point dipole model by a quadrupole tensor of
symmetry and strength appropriate to
reasonably represent the water molecule. As a result of the
highly orientational interactions between these water-like molecules an
ice-like solvent structure is formed near a smooth, curved \cite{tor88} and
planar surface \cite{kin94} in a region, which is usually attributed to
the 'inner layer' of the double layer structure.
Adding an octupole moment to the model yields a net polarization of the
interfacial solvent structure at the neutral surface \cite{tor93}. In spite
of some difficulty in making the electrode completely planar, the interfacial
structure obtained with these multipolar water models resembles computer
simulation results for more elaborate water models \cite{lee84}.

These more sophisticated steric water models (e.g. ST2, TIP4P and SPC)
have been employed mainly in computer simulations because of their
complexity. Several authors have focused their attention on how the detailed
atomic structure of the electrode affects the interfacial structure of pure
water for instance through chemical bonding, surface corrugation
\cite{hei89,rag91} and image interactions modelling a metal electrode
\cite{hau89}. The model parameters are based on quantum mechanical ab initio
calculations or on thermodynamic data. Present day computer simulations are
restricted, especially for an aqueous electrolyte near an electrode, to
systems with only a few ions and a few hundred of water molecules
representing a highly concentrated electrolyte. Serious problems arise also
from limitations in computing time, because long simulation runs are
necessary to observe the ions entering and leaving the interfacial region
\cite{glo92,ros91}, and special molecular dynamics techniques have been
applied \cite{spo93}. Therefore it is difficult to obtain statistically
reliable results from computer simulations near an interface. However,
computer simulations for the complex electrode-electrolyte interface are
just at the beginning. Recently, first results of an X-ray diffraction
experiment indicated some interesting water structure near the electrode
as well \cite{ton94}.

In comparision with the simulations, integral equation methods are more
versatile when considering less concentrated electrolytes or when applying an
external electric field on the interfacial structure and varying the field
strength. This is in particular important for calculating the differential
capacitance. In this paper we use the Wertheim-Lovett-Mou-Buff (WLMB)
integral equation for determining the density profiles at the
uncharged and charged electrode. We take the central force model as an
appropriate water model for our calculations. The most dramatic effect of the
specific water structure established near a planar surface on the
electrical double layer is to be expected for ions with small cationic and
large anionic size. Such a candidate is sodium chloride dissolved in water,
which is particularly important in many electrochemical and biological
applications.

In Sec. II. the model for the aqueous solution of sodium chloride and the
method for calculating bulk correlation functions is described. Results for
the structure of the ionic hydration shell and for ion pairing are presented
and discussed. In the first part of Sec. III.\, the WLMB-equation for the
density profiles of a mixture of charged particles in contact with a planar
electrode is given. The equation requires particle-particle correlations,
which are taken from the bulk fluid. In the region near the electrode the
water solvent shows the ice-like layering. We discuss the effect
of the ice-like water structure on the distribution of the
differently sized sodium and cloride ions. In addition, we study the
influence of an external electric field on the interfacial structure and the
implications on the differential capacitance. Conclusions are summarized in
Sec. IV.

\section{The bulk electrolyte}
\noindent
{\bf A. Model and computational method.} The model electrolyte used within
this paper is designed to represent sodium chloride dissolved in water.
As an appropriate water model for the integral equations we employ
the central force (CF) water model in the revised version \cite{sti78}.
The central force model describes water as a
stoichiometric mixture of partially charged hydrogen and oxygen atoms
($q_{H}=0.32983 \, e$ and $q_{O}=-0.65966 \, e$) interacting via pairwise
additive central force potentials, which are designed to
reproduce the formation of water molecules and their tetrahedral coordination
via hydrogen bonds. The detailed form of the pair potentials are given in
reference \cite{vos94}. Implicitly, the water molecules formed by the central
force pair potentials are flexible. In spite of its simplicity this water
model is known to yield good results for the bulk structure and for the water
structure at a hard wall \cite{vos94}.

The interaction potential between the ion and each atom of the water molecule
consists of a Lennard-Jones pair potential added to the Coulomb interaction
\begin{equation}
\label{lj}
u_{\alpha\beta}(r) = \frac{q_{\alpha} q_{\beta} }{r} \; + \; 4 \,
\epsilon_{\alpha\beta} \;
\left[ {\left( \frac{ \sigma_{\alpha\beta} }{r} \right)}^{12} \; - \;
{\left( \frac{ \sigma_{\alpha\beta} }{r} \right)}^{6} \right].
\end{equation}
We have fitted, according to the procedure proposed by Pettitt et al.
\cite{pet86}, the Lennard-Jones parameters to the Hartree-Fock binding energy
of the ion-water dimer in the equilibrium configuration, which has been
determined for some alkali- and halide-water dimers by Kistenmacher et al.
\cite{kis73} with quantum mechanical ab initio calculations. The
Lennard-Jones parameters for the interaction between a sodium or chloride
ion and the atoms of a CF-water molecule are listed in table \ref{tab1}.
This parametrization of the ion-water interaction potential is by no means
unique. Another set of pair potentials for describing the interaction between
a sodium or chloride ion and the CF water molecule were derived by Bopp et
al. \cite{bop79}. Their pair potentials are able to yield a quite
satisfactory representation of the dimer energy surface obtained from the
ab initio calculations over a wide range of ion-water seperation.
But they imply some important inconsistencies in the asymptotic behaviour of
the correlation functions and as a result some thermodynamic properties, for
instance the pressure calculated according to the virial equation or the
isothermal compressibility, are not defined.

The pair potentials between like ions are of the Huggins-Mayer form
\begin{equation}
\label{mh}
u_{\alpha\alpha}(r) = \frac{q_{\alpha}^{2}}{r} \; + \;
B_{\alpha\alpha} \, \exp\/[- \, r / \rho_{\alpha\alpha}] \; - \;
\frac{C_{\alpha\alpha}}{r^{6}}
\end{equation}
with potential parameters taken from \cite{pet86}.  The interaction between
the sodium and the chloride ion was modelled in addition to the Coulomb
interaction by a Lennard-Jones pair potential with parameters
$\epsilon_{NaCl}=2.816 \times 10^{-14}erg$ and $\sigma_{NaCl}=3.6\, \AA$
as proposed by Smith and Haymet. \cite{smi92}. Taking instead the
Huggins-Mayer form of reference \cite{pet86} for the pair interaction between
the unlike ions and considering an ion concentration of 0.01 molar,
convergent solutions of the Ornstein-Zernike equation are not obtained with
the iteration procedure used for calculating the bulk correlation functions.
We have found that convergence is not achieved because of a thermodynamic
instability of the electrolyte solution, which can be characterized as a
demixing of ions and the water solvent. The demixing cannot be avoided by
changing the form of the interaction potential or the approximations employed
in the iteration scheme, only the onset can be shifted \cite{vos95}.

Since within the chosen model electrolyte the aqueous solution of sodium
chloride is simply a homogeneous mixture of charged particles interacting
via additional short ranged, spherically symmetric pair interactions,
the Ornstein-Zernike equation for calculating the correlation functions reads
\begin{equation}
\label{oz}
h_{\alpha\/\beta}\/(r) = c_{\alpha\/\beta}\/(r) + \sum_{\gamma} \rho_{\gamma}
\int \, d\/{\bf r}^{'} \: h_{\alpha\/\gamma}\/({r}^{'}) \,
c_{\gamma\/\beta}\/(|{\bf r} - {\bf r}^{'}|)
\end{equation}
together with the closure relation
\begin{equation}
\label{clo}
\ln\/g_{\alpha\/\beta}\/(r) = - \frac{1}{k_{B}T} u_{\alpha\/\beta}\/(r)
+ h_{\alpha\/\beta}\/(r) - c_{\alpha\/\beta}\/(r) - B_{\alpha\/\beta}\/(r).
\end{equation}
$k_{B}$ is the Boltzmann constant, $T$ is the temperature and
$\rho_{\gamma}$ denotes the number
density of particle species $\gamma$. Including the so called bridgefunction
$B_{\alpha\/\beta}\/(r)$ in the closure equation (\ref{clo}) gives an exact
relation between the pair potential $u_{\alpha\beta}$ and the pair
distribution function $g_{\alpha\beta}=1+h_{\alpha\beta}$. These
bridgefunctions are known only for simple liquids and mixtures but not for
more complex fluids like water or the aqueous electrolyte. Therefore these
functions have to be approximated. $B_{\alpha\beta}=0$ is the familiar
hypernetted chain (HNC) approximation. Application of the
HNC-approximation for calculating the water correlation functions failed to
reproduce the structural properties of pure water \cite{thu83}. Previously
\cite{vos94}, we have developed bridgefunctions for the central force water
model quite similar to those proposed by Ichiye et al. \cite{ich88}.
Employing our bridgefunctions in the Ornstein-Zernike formalism yields a
faithfull description of the structural and thermodynamic properties of pure
water under standard conditions. For instance two hydrogen atoms are bound to
an oxygen atom forming together the triangular water molecule. Approximately
two more hydrogen atoms are located near the oxygen atom at a distance, which
is comparable to the hydrogen bond length indicating the formation of the
hydrogen bonded water network.
We have added these bridgefunctions in the closure relation (\ref{clo}) for
the water correlations in the electrolyte, while the remaining ion-water and
ion-ion correlation functions are calculated with the HNC-approximation.
The set of coupled Ornstein-Zernike integral equations (\ref{oz}) together
with the closure relations (\ref{clo}) are solved iteratively at fixed
temperature T, water density and ion concentration. The results are the
correlation functions $h_{\alpha\beta}(r)$ and $c_{\alpha\beta}(r)$ in the
bulk electrolyte. For details of the numerical calculation the reader is
refered to reference \cite{vos94}. In the calculations, presented in this
article, the temperature is fixed at $T = 300 \, K$ and the number densities
for hydrogen and oxygen atoms are chosen
$\rho_{H} = 0.06690 \, {\AA}^{-3}$ respectively
$\rho_{O} = 0.03345 \, {\AA}^{-3}$ equivalent to the mass density $\rho = 1 \,
g / cm^{3}$. \\*

\noindent
{\bf B. Structural results.} With the integral equation method presented above
we have calculated bulk correlation functions for a 0.01 molar aqueous
solution of sodium chloride, which means 5555 water molcules per sodium
chloride molecule. In Fig.\ref{fig1}a and \ref{fig2}a the pair distribution
functions $g_{\alpha\beta}(r)$ of hydrogen and oxygen atoms around a central
$Na^{+}$- and $Cl^{-}$-ion, respectively, are shown. Both ion-oxygen
distribution functions exhibit a sharp first maximum determining the position
of water molecules in the first hydration shell around the ion. We found the
first oxygen atoms at a distance $r_{NaO}=2.34 \AA$ apart from the sodium ion
and at $r_{ClO}=3.52 \AA$ for the chloride ion. The positions of hydrogen and
oxygen atoms in the ionic hydration shells are summarized in table \ref{tab2}.

Due to the stronger ion-water interaction the hydration shell of the smaller
sodium ion is more pronounced. The broad maximum of the function
$g_{NaH}(r)$ is obviously a superposition of two seperate maxima with
different weight. We have fitted two gaussian functions to this part of the
curve to determine the position of the two maxima. The first maximum is found
at $r_{NaH}=2.69 \AA$ and the second at $r_{NaH}=3.21 \AA$. The hydrogen
atoms contributing to the first hydration shell of the $Na^{+}$-ion are
located at distances slightly smaller and larger than expected for a single
water dipole perfectly aligned in the radial electric field of the sodium
ion. The arrangement of atoms described by the pair distribution functions
can be interpreted as the water dipole in the $Na^{+}$-hydration shell tilted
with respect to the oxygen-ion vector and one of the 'lone pair'-valencies of
the water molecule is orientated towards the ion. Obviously, this orientation
is energetically more favourable because one more hydrogen bond with water
molecules in the second hydration shell can be established compared to the
configuration where the water dipole is radialy pointing away from the sodium
ion and always both 'lone pair'-valencies are directed towards the ion. Due
to the packing of water molecules in the first hydration shell some molecules
are also tilted with respect to the plane formed by the dipole vector and the
oxygen-ion position vector. This leads to the double maximum of $g_{NaH}(r)$.
In order to evaluate the mean orientation of water molecules in the cationic
hydration shell and to compare with computer simulation results it is
convenient to consider the angle $\theta$ enclosed by the dipole vector of the
water molecule and the oxygen-ion position vector. We deduce the mean value
of $\theta$ from the positions of the maxima of the pair distribution
functions $g_{\alpha\beta}(r)$ under the assumption that the geometry of
the water molecule in the hydration shell is not strongly perturbed by the
neighbouring ion. This assumption seems to be well justified by computer
simulations \cite{gua90}. A value of $\theta=100^{o}$ corresponds to the
first maximum of $g_{NaH}(r)$ and $\theta=145^{o}$ to the second. With the
weight of the gaussian fit functions we obtain $\theta=117^{o}$ for the mean
orientation of water molecules in the cationic hydration shell. Our results
for $\theta$ are summarized in table \ref{tab3} together with results
obtained in computer simulations. With regard to the results of Heinzinger et
al. \cite{bop79}, $\theta$ is obtained from their values of
$\langle \cos\/\theta \rangle$. The second set of values for the mean
orientation is deduced from the pair distribution functions reported by
Smith et al. \cite{smi92}. Some uncertainties should be attributed to the
results obtained from the pair distribution functions, since the angle
$\theta$ is very sensitive to the geometrical parameters of the water model
and to the location of pair distribution maxima. Nethertheless,
$\theta$ calculated with the integral equation method is in fairly good
agreement with the computer simulation results.

In addition we have calculated the running coordination
numbers
\begin{equation}
N_{\alpha\/\beta}\/(r) = 4 \pi \, \rho_{\beta} \, \int_{0}^{r} \:
dx \, x^{2} \, g_{\alpha\/\beta}\/(x)
\end{equation}
of hydrogen and oxygen atoms around the central ion in order to characterize
the hydration shell. In Fig.\ref{fig1}b and \ref{fig2}b the coordination
numbers are plotted as a function of distance from the center of the ion. As a
consequence of the pronounced $Na^{+}$-hydration shell the function
$N_{NaO}(r)$ possesses a well defined plateau. The coordination numbers,
listed in table \ref{tab2}, are defined as the value of $N_{\alpha\beta}(r)$
at the first minimum of $g_{\alpha\beta}(r)$. Around the $Na^{+}$-ion we
found on average 6.5 oxygen atoms. In our calculation the ratio
$N_{NaH}/N_{NaO}$ of the coordination numbers according to the first minima
of the pair distribution functions is 2.28, quite similar to a simulation
result \cite{gua90}, indicating that also hydrogen atoms of the second
hydration shell contribute to the first $g_{NaH}(r)$-maximum.

In a dilute solution the energy of ionic solvation consists of two parts:
the solvent reorganization energy due to the creation of a polarized cavity
in the pure solvent and the interaction energy between the ion and the
surrounding water molecules already properly reordered. In the present
work the ion-solvent contribution to the solvation energy is calculated
by determining the potential energy of the ion in its hydration shell
according to
\begin{equation}
\label{usolv}
U_{\alpha}^{solv} = 4 \pi \, \sum_{\beta=O,H} \rho_{\beta} \,
\int_{0}^{\infty} d\/r \, r^{2} \, g_{\alpha\beta}(r) \, u_{\alpha\beta}(r).
\end{equation}
The results for the $Na^{+}$- and the $Cl^{-}$-ion are presented in table
\ref{tab4}. $U_{\alpha}^{solv}$ is dominated by the Coulomb potential close
to the ion. The orientation of water dipoles around the smaller $Na^{+}$-ion
is more pronounced due to the strong electric field and therefore
$U_{\alpha}^{solv}$ is more negative for the $Na^{+}$-ion. Our value
$U_{Na^{+}}^{solv} = - 8.60 \, eV$ is close to the result of Chandrasekhar
et al. \cite{cha84} for the rigid TIPS-water model.

The total work required to bring an additional ion into the system is given
by the chemical potential $\mu_{\alpha}$. The excess part
$\mu_{\alpha}^{ex}$ due to particle interactions can be determined from the
two-particle correlation functions \cite{che91} and in case of the
HNC-approximation this relation reads
\begin{equation}
\mu_{\alpha}^{ex} = 2 \pi \, \sum_{\beta=H,O,ion} \rho_{\beta}
\int_{0}^{\infty} d\/r \, r^{2} \, \left[ h_{\alpha\beta}^{2}(r) -
g_{\alpha\beta}(r) \, c_{\alpha\beta}(r) - c_{\alpha\beta}(r) \right].
\end{equation}
In addition to the potential energy gain $U_{\alpha}^{solv}$ of the ion due
to ion-water interactions the excess part $\mu_{\alpha}^{ex}$ of the chemical
potential includes also the work required for creating a cavity for the ion in
the water network and therefore $\mu_{\alpha}^{ex}$ must be less negative than
$U_{\alpha}^{solv}$. Our value $\mu_{Na^{+}}^{ex} = -3.23 \, eV$ follows
this prediction and is in good agreement with the value $-3.93 \, eV$
determined experimentally for the free energy of solvation of a
$Na^{+}$-ion \cite{gom77}.

Around the $Cl^{-}$-ion (cf. Fig.\ref{fig2}a) hydrogen atoms are arranged in
two distinctly seperated shells and the corresponding oxygen atoms are
located in between. This distribution of atoms can be interpreted in terms of
a water configuration in the anionic hydration shell where one hydrogen atom
of the water molecule is directed into the liquid while the second is
pointing towards the $Cl^{-}$-ion. Similar structural results were observed
in analogous computer simulation studies \cite{bop79,smi92} and in neutron
diffraction experiments \cite{end87}. Due to such almost linear orientation
of one oxygen-hydrogen bond towards the chloride ion
each water molecule in the first $Cl^{-}$-hydration shell should contribute
only with one hydrogen and one oxygen atom to the first $g_{ClH}(r)$- and
$g_{ClO}(r)$-maximum. This is expressed by the calculated coordination
numbers $N_{ClH}=18.2$ and $N_{ClO}=17.9$, although in comparision with the
computer simulation result \cite{gua90} they are too large. This discrepancy
may be attributed to the chosen ion-water pair interactions and to the
HNC-approximation employed in the integral equation procedure. The
coordination numbers are very sensitive on the details of the pair
distribution functions, especially on the position and the magnitude of both
the first $g_{\alpha\beta}(r)$-maximum and -minimum. As the size of the ion
increases the coordination numbers increases too because of an increasing
volume available for water molecules in the first hydration shell. On the
other hand the resolution of the first hydration shell decreases, as
exemplified by the weak coordination plateau in $Cl^{-}$-hydration shell
presented in Fig.\ref{fig2}b, because of the weaker electrostatic attraction.
As a result of the overpopulation of the anionic hydration shell, we obtain
rather large values of $U_{Cl^{-}}^{solv} = - 7.34 \, eV$
and $\mu_{Cl^{-}}^{ex} = - 6.25 \, eV$ compared with computer simulation
results (cf. table \ref{tab4}) and the experimental value of $- 3.53 \, eV$
for the free energy of solvation \cite{gom77}. With the RISM integral
equation method Pettitt et al. \cite{pet86} calculated for the $Cl^{-}$-ion
a more negative ion-water contribution to the solvation energy than for the
$Na^{+}$-ion. This is clearly in disagreement with several computer
simulations and contradicts the idea, that the solvation shell of the smaller
cation should be bound more strongly because of the larger electrostatic
attraction. The too short chloride-hydrogen distance might be the reason for
their very large value of $U_{Cl^{-}}^{solv}$.

The calculated structural properties of the ionic hydration shells for a
dilute aqueous solution of sodium chloride are summarized in table \ref{tab2}
and compared with available experimental data obtained by x-ray \cite{new89}
or neutron diffraction techniques \cite{nei85} and with computer simulation
\cite{bop79,smi92} and other integral equation results
\cite{pet86,thu83,ich88}. The results for the position of the first
ion-oxygen and ion-hydrogen peaks are in good agreement with the experimental
data and with those obtained by Smith et al. using the same CF-water model but
different ion-water pair potentials. In the calculation of Thuraisingham et
al. all bulk correlation functions, in particular for CF-water, are obtained
with the HNC-approximation. This approximation does not reproduce the
structural features of bulk water. As a result the number of oxygen atoms in
the first hydration shell is overestimated. An improvement is achieved in the
present work by adding appropriate bridgefunctions in the closure relation
for the water correlation functions. Also some results of an integral
equation theory (RISM) for interaction site fluid models are added in
table \ref{tab2} \cite{pet86}.

Since in a dilute ionic solution only a few water molecules are interacting
directly with the ion, the pair distribution functions $g_{HH}$, $g_{OH}$
and $g_{OO}$ are indistinguishable from those of pure water reported in
\cite{vos94}.

The averaged structural effects of the solvent on the ion-ion interactions
may be discussed in terms of the potential of mean force (PMF), defined as
\begin{equation}
W_{\alpha\beta}(r) = -k_{B} T \, \ln\/g_{\alpha\beta}(r).
  \label{pmf}
\end{equation}
Our results for the potential of mean force (cf. Fig.\ref{fig3} and
Fig.\ref{fig4}) show some characteristic structure due to the molecular
nature of the solvent. In comparision with the bare pair interaction
potential $u_{\alpha\beta}(r)$, plotted in Fig.\ref{fig3}b, the first minimum
of $W_{NaCl}(r)$ is drastically reduced in magnitude and slightly shifted
towards larger r-values. This strongly reduced attraction between the
$Na^{+}$- and $Cl^{-}$-ion at contact arises primarily from a partial
overlap of the hydration shells of the single ions. Debye-H\"uckel screening
is of minor importance. The first minimum of $W_{NaCl}(r)$ at $r=3.42 \AA$
represents the 'contact ion pair' (CIP), whereas the second minimum at
$r=5.36 \AA$ corresponds to the 'solvent seperated ion pair' (SSIP),
suggesting a solvation bridge between the sodium and the chloride ion.
A combination of the configurations depicted in Fig.\ref{fig1}a and
Fig.\ref{fig2}a for one bridging water molecule explains the position of the
SSIP-minimum. For the intervening water molecule a position is energetically
more favourable, where the orientation matches those in the hydration
shell of a single $Na^{+}$- and $Cl^{-}$-ion. Therefore the water molecule
is slightly displaced from the line connecting both ion centers and the
hydrogen atoms are orientated towards the $Cl^{-}$-ion. This
interpretation is in line with a previous molecular dynamics simulation
\cite{gua91}, where the solvent configuration due to the minima of
$W_{NaCl}(r)$ is inspected.

The relative depth of the minima determines the stability of the CIP- and the
SSIP-state. In our calculation with the CF-water model the first minimum of
$W_{NaCl}(r)$ is found to be deeper than the second. The barrier for ionic
dissociation (CIP $\rightarrow$ SSIP) is $0.97 \, kcal/mol$ and for ionic
association (SSIP $\rightarrow$ CIP) $0.30 \, kcal/mol$. The barrier
for the dissociation process is higher than for the association process and
the CIP state is the most stable configuration. An indication whether the
associated or the dissociated state is prefered by the sodium-chloride ion
pair is provided by the ionic association constant \cite{jus76}
\begin{equation}
K_{a} = 4 \pi \, N_{A} \, \int_{0}^{R} d\/r \, r^{2} \,
\exp\/\left[ - \; \frac{W_{NaCl}(r)}{k_{B}T} \right],
\label{asconst}
\end{equation}
where $N_{A}$ denotes the Avogadro constant and $R$ is the position of the
barrier maximum seperating the SSIP- from the CIP-state. For the 0.01 molar
solution of sodium chloride we obtain $K_{a} = 10.13 \, l/mol$. In a dilute
ionic solution with concentration c the coordination number $K_{a} \times c$
is interpreted as the fraction $\theta$ of unlike CIP-pairs in the system.
Recent MD simulations suggest the tendency for ion pairing to increase
when the dipole moment of the water molecule decreases \cite{gua91}. Compared
to the dipole moments of the water models employed in these simulations the
mean dipole moment of the flexible CF-water molecule in the fluid phase
($\langle \mu \rangle = 2.154 \, D$), which is deduced from our pair
distribution functions calculated with the HNC+B-approximation, is smaller.
Our larger value for $K_{a}$, when compared with the simulation results, is
therefore consistent with the higher probability of the CIP configuration
when water models with a smaller dipole moment are employed.

Despite the purely repulsive nature of the pair interaction between like
ions, the respective potentials of mean force, as plotted in Fig.\ref{fig4},
show some oscillatory structure. The existence of minima in the
PMFs indicates ion pairing resulting from a subtle balance between the
interionic repulsion and an effective attraction mediated by the ionic
solvation shells. The PMF for $Na^{+}-Na^{+}$ and for $Cl^{-}-Cl^{-}$ are
significantly different. The tendency for solvent-bridged sodium ion pairing
is rather weak as indicated by the shallow minimum of $W_{NaNa}(r)$ at
$r=4.52 \AA$ and around $6 \AA$. These positions are in fairly good agreement
with a previous computer simulation using the CF-model \cite{smi92}. The
first minimum of $W_{NaNa}(r)$ corresponds to a configuration where a water
molecule is in a bridge position between the two sodium ions with the
hydrogen atoms pointing into the liquid \cite{gua91a}. In contrast to the
$W_{NaNa}(r)$ our calculated $W_{ClCl}(r)$ possesses a prominent stable
minimum at $r=6.6 \AA$. With the characteristic bond lengths in the
chloride-water dimer we found this position consistent with a configuration
where a water molecule is placed between the chloride ions orientating each
of its hydrogen atoms towards one of the ions. The water molecule serves as a
bridge in the chloride ion pair. The relative stability of the solvent-bridged
chloride ion pair is known from neutron scattering experiments on highly
concentrated electrolyte solutions. The $Cl^{-}-Cl^{-}$ pair distribution
function extracted from the measured partial structure factors shows a
prominent peak at $r=6.1 \AA$ for the aqueous solution of $Ni_{2}Cl$
\cite{nei83} and at $r=6.4 \AA$ for the $LiCl$-solution \cite{cop85}. In
addition to the 'dilute peak' around $6 \AA$, a second maximum of
$g_{ClCl}(r)$ is observed in neutron scattering experiments at $r=3.75 \AA$
only for high ionic concentration. This peak is reminiscent of the molten
salt regime and dimishes as the solution becomes more dilute. The RISM
integral equation approach predicts only one stable minimum of the
$Cl^{-}-Cl^{-}$ PMF at $r=3.5 \AA$ even at infinite dilution and obviously
fails to reproduce the 'dilute peak' \cite{pet86}. Although the positions of
stable minima of the PMF are solely determined by the specific ion-water
interactions, we found that the depth of the minima and therefore the
stability of the ion pairs depend strongly on the ion concentration and the
approximations employed in the integral equation calculations.

\section{The electrolyte at the electrode}
\noindent
{\bf A. Computational method.} Several integral equations for calculating
the density profiles at an electrode-electrolyte interface exist. A first
relation between the particle density $\rho_{\alpha}(1)$ and the potential
energy $V_{\alpha}(1)$ of a particle at coordinate 1 due to an external
potential, e.g. the electrode, is given by the Born-Green-Yvon (BGY)
equation \cite{han86}
\begin{equation}
\label{bgy}
\nabla_{1} \ln\/\rho_{\alpha}(1) = - \, \frac{1}{k_{B}T} \left[ \nabla_{1}
V_{\alpha}(1) + \sum_{\beta} \int d\/2 \, \rho_{\beta}(2) \,
g_{\alpha\beta}(1,2;[\rho]) \, \nabla_{1} u_{\alpha\beta}(1,2) \right],
\end{equation}
where $u_{\alpha\beta}(1,2)$ is the pair potential and $g_{\alpha\beta}(1,2;
[\rho])$ denotes the pair distribution function in the inhomogeneous density
distribution in front of the electrode. The functional dependence of
$g_{\alpha\beta}$ on the whole set of one-particle densities
$\{\rho_{\gamma}(1)\}$ is indicated by the notation
$g_{\alpha\beta}(...;[\rho])$. By density functional arguments
\cite{eva92,alt87} an alternative to the BGY-equation can be
derived, where the density profile is related to the direct correlation
function $c_{\alpha\beta}$ instead of the total correlation function
$h_{\alpha\beta}$
\begin{equation}
\label{hnc}
\ln \left[ \frac{\rho_{\alpha}(1)}{\rho_{\alpha}} \right] = - \,
\frac{1}{k_{B}T} \, V_{\alpha}(1) + \sum_{\beta} \int d\/2 \,
[\rho_{\beta}(2) - \rho_{\beta}] \, \int_{0}^{1} d\/\lambda \,
c_{\alpha\beta}(1,2;[\rho_{\lambda}]).
\end{equation}
This equation is exact too, but requires a functional integration
of the direct correlation function over a sequence of density distributions
$[\rho_{\lambda}]=\{ \rho_{\gamma}(1;\lambda) \}$ from the initial state
($\lambda=0$), e.g. the homogeneous fluid, to the final state ($\lambda=1$),
when the external potential is completely switched on. By the same density
functional arguments an even simpler relation between the density gradient
and the external potential can be obtained, which avoids the functional
integration. Originally this equation was derived by Wertheim \cite{wer76}
and independently by Lovett, Mou and Buff \cite{lov76} for simple fluids,
hereafter called the WLMB-equation. In the generalisation for a particle
mixture the WLMB-equation reads
\begin{equation}
\label{wlmb}
\nabla_{1} \ln\/\rho_{\alpha}(1) = - \, \frac{1}{k_{B}T} \, \nabla_{1}
V_{\alpha}(1) + \sum_{\beta} \int d\/2 \, c_{\alpha\beta}(1,2;[\rho]) \,
\nabla_{2} \rho_{\beta}(2).
\end{equation}
In contrast to eq.(\ref{hnc}) the BGY- and WLMB-equation require only the
inhomogeneous particle-particle correlations in the final density
distribution near the electrode and not for a sequence $[ \rho_{\lambda} ]$
of densities. Calculating the inhomogeneous particle-particle correlation
functions from the Ornstein-Zernike equation requires great computational
effort. Only for much simpler systems like the primitive model electrolyte in
contact with a hard wall, this complicated calculation has been done
\cite{kje92}. For this reason several approximations for the inhomogeneous
particle-particle correlations have been applied in the density profile
equations (\ref{bgy})-(\ref{wlmb}). Our experience has lead us to the
conclusion that an approximation of the direct correlation function
$c_{\alpha\beta}(1,2;[\rho])$ is more robust than of the total correlation
function $h_{\alpha\beta}(1,2;[\rho])$ \cite{nie85,kas93}. In the simplest
approximation the inhomogeneous direct correlation function is replaced by
its uniform-fluid counterpart
\begin{equation}
\label{chom}
c_{\alpha\beta}(1,2;[\rho]) \approx
c_{\alpha\beta}(1,2; \{ \rho_{\gamma} \}) = c_{\alpha\beta}(1,2).
\end{equation}
Going a step further, the inhomogeneous direct correlation functions can be
locally approximated by correlation functions of the homogeneous fluid
evaluated at appropriate weighted densities \cite{eva92,nie85,kas93,kas92}.
For our complex model electrolyte we have achieved very good results using the
approximation (\ref{chom}). Especially the ice-like water layers in
the vincinity of the electrode were formed \cite{vos94}. On this stage of
approximation eq.(\ref{hnc}) and the WLMB-equation (\ref{wlmb}) become
equivalent. For numerical application we prefer the WLMB-equation.

Considering a planar electrode perpendicular to the z-axis  of a Cartesian
coordinate system, the one-particle density $\rho_{\alpha}(1)$ and the
external
potential energy $V_{\alpha}(1)$ are functions of z only. The gradient is
a derivative with respect to z and the integration perpendicular to the z-axis
can be carried out seperately. This leads to the WLMB-equation in the form
\begin{equation}
\label{wlmb1}
\frac{ d\/\ln\/\rho_{\alpha}(z_{1}) }{ d\/z_{1} } = - \frac{1}{k_{B}T} \,
\frac{ d\/V_{\alpha}(z_{1}) }{ d\/z_{1}} + 2 \pi \, \sum_{\beta}
\int_{-\infty}^{\infty} d\/z_{2} \,
\frac{ d\/\rho_{\beta}(z_{2}) }{ d\/z_{2} }  \,
\int_{|z_{1}-z_{2}|}^{\infty} d\/r \, r \, c_{\alpha\beta}(r).
\end{equation}
Since our model electrolyte consists of charged particles, eq.(\ref{wlmb1})
contains diverging terms due to the Coulombic interactions. For an
analytical treatment of these long-range terms seperating the short-range
part and the Coloumbic tail of the direct correlation function, as proposed by
Ng \cite{ng74}, is appropriate
\begin{equation}
c_{\alpha\beta}(r) = c_{\alpha\beta}^{SR}(r) - \frac{ q_{\alpha}
q_{\beta} }{k_{B}T} \, \frac{ erf\/(\lambda r)}{r},
\end{equation}
where $erf\/(\lambda r)$ is the error-function. With the definition of the
charge density $q(z)=\sum_{\alpha} q_{\alpha} \rho_{\alpha}(z)$ the diverging
Coulombic terms are collected under the sum over $\beta$. Requiring bulk
neutrality ($\rho_{\alpha}(z) \stackrel{z \rightarrow \infty}{=}
\rho_{\alpha}$) and $\rho_{\alpha}(z \leq 0)=0$ in the left halfspace,
the charge density vanishes on the boundaries of the integral and integration
by parts yields
\begin{eqnarray}
\label{wlmb2}
\frac{ d\/\ln\/\rho_{\alpha}(z_{1}) }{ d\/z_{1} } & = & - \frac{1}{k_{B}T} \,
\frac{ d\/V_{\alpha}(z_{1}) }{ d\/z_{1}} + 2 \pi \, \sum_{\beta}
\int_{-\infty}^{\infty} d\/z_{2} \,
\frac{ d\/\rho_{\beta}(z_{2}) }{ d\/z_{2} } \int_{|z_{1}-z_{2}|}^{\infty}
d\/r \, r \, c_{\alpha\beta}^{SR}(r) \nonumber \\
 &   & + \, \frac{ 2 \pi q_{\alpha} }{k_{B}T} \,
\int_{-\infty}^{\infty} d\/z_{2} \, q(z_{2}) \, sgn(z_{1}-z_{2}) \,
erf(\lambda |z_{1}-z_{2}|).
\end{eqnarray}
In order to investigate the influence of an external electric field
on the structural properties of the electrode-electrolyte interface, the
elctrode is covered with a homogeneous charge density $\omega$:
\begin{equation}
\label{pot}
V_{\alpha}(z_{1}) = V_{\alpha}^{SR}(z_{1}) - 2 \pi q_{\alpha} \omega z_{1}.
\end{equation}
The repulsive nature of the electrode or some specific chemical bonding is
incorporated in the short ranged interaction potential
$V_{\alpha}^{SR}(z_{1})$. Under the constraint of exact compensation of the
surface charge density by charges in the fluid, the mean electric field due
to the charge distribution in the right halfspace ($z_{1} \geq 0$)
\begin{equation}
\label{efield}
E(z_{1}) = - \, 4 \pi \, \int_{z_{1}}^{\infty} d\/z_{2} \, q\/(z_{2}).
\end{equation}
satisfies the boundary condition $E(z_{1}=0)=4 \pi \omega$ at the electrode.
With this assumption, using eq.(\ref{pot}) and eq.(\ref{efield}) and
introducing the complementary error-function $erfc(x)=1-erf(x)$, equation
(\ref{wlmb2}) can be rewritten in the form
\begin{eqnarray}
\label{wlmb3}
\frac{ d\/\ln\/\rho_{\alpha}(z_{1}) }{ d\/z_{1} } & = & - \frac{1}{k_{B}T} \,
\frac{ d\/V_{\alpha}(z_{1}) }{ d\/z_{1}} +
\frac{q_{\alpha} E(z_{1})}{k_{B}T} + 2 \pi \sum_{\beta}
\int_{-\infty}^{\infty} d\/z_{2} \,
\frac{ d\/\rho_{\beta}(z_{2}) }{ d\/z_{2} } \int_{|z_{1}-z_{2}|}^{\infty}
d\/r \, r \, c_{\alpha\beta}^{SR}(r) \nonumber \\
 & - & \frac{ 2 \pi q_{\alpha} }{k_{B}T} \,
\int_{-\infty}^{\infty} d\/z_{2} \, q(z_{2}) \, sgn(z_{1}-z_{2}) \,
erfc(\lambda |z_{1}-z_{2}|).
\end{eqnarray}
The information about particle-particle correlations are contained in the
short ranged part of the direct correlation functions, calculated seperately
from the Ornstein-Zernike equation for the homogeneous mixture as discussed
in section {\rm II}. Starting with bulk density $\rho_{\alpha}$ at
$z_{1} \rightarrow \infty$ an integration yields the density profile
\begin{equation}
\rho_{\alpha}(z_{1}) = \rho_{\alpha} - \int_{z_{1}}^{\infty} \, d\/z_{2} \,
\frac{ d\/\rho_{\alpha}(z_{2}) }{d\/z_{2} }.
\end{equation}
The set of coupled WLMB-equations (\ref{wlmb3}) was solved iteratively
applying a standard mixing scheme to stabilize the iteration procedure.
In each iteration step the condition of charge neutrality has been carefully
controlled. With this method the density profiles for an aqueous solution of
sodium chloride in contact with an uncharged and charged electrode are
calculated. The electrode was assumed unpolarizable. The short-range
interaction $V_{\alpha}^{SR}(z)=A/z^{9}$ ($A=1.88 \times 10^{-14} erg$)
of the electrode is purely repulsive and equal for all particles. This fast
decaying repulsive potential is very similar to a hard wall, but convergence
of the iteration scheme is achieved more rapidly for the 'soft' wall. \\*

\noindent
{\bf B. Results for the uncharged electrode. }Fig.\ref{fig5}a shows the
density profiles for sodium and chloride ions as well as for the hydrogen and
oxygen atoms calculated for the 0.01 molar solution at the uncharged,
repulsive electrode. In front of the electrode there is a densely packed
water layer. About $3 \AA$ apart a second less densely packed
layer is found. This distribution of hydrogen and oxygen atoms resembles
the density profiles obtained for pure water at the same type of wall,
which have been interpreted in terms of an ice-like water structure
\cite{vos94}. In comparision to these previous calculations the density of
the first water layer is slightly reduced. Both ion species are strongly
repelled by this densely packed water layer. In the region of the second
water layer sodium ions are found with almost the same probability as
in the bulk, whereas the larger chloride ions are still hindered from
penetrating into this part of the interfacial region. Since the electrode
acts equally repulsive on both ion species, steric and electrostatic
interactions between the differently sized ions and the ice-like water matrix
are responseable for the difference in the ionic density profiles.

Information about the water structure is obtained by inspecting the
charge density $q(z)=q [\rho_{H}(z) - 2 \rho_{O}(z)]$. In Fig.\ref{fig5}b
this difference of particle densities is plotted for the water solvent
in the electrolyte and for pure water. Both charge density profiles
are very similar; that means the water structure at the electrode is not
much perturbed by the presence of ions. The sequence of positive and negative
charge density is characteristic for the ice structure. An schematic
illustration of the ice structure can be found for example in reference
\cite{vos94}. Each water molecule provides two hydrogen atoms and two
'lone pair'-valencies in tetrahedral geometry. Since reorientation of water
molecules without disrupting the hydrogen bonded network is possible under
the constraint that each hydrogen bond is occupied by only one hydrogen atom,
the ice structure has some internal polarizability. Thus the ice-like water
structure can persist also under the influence of external electric fields.
The charge
density profile of water at a planar electrode has been already disscused in
detail previously \cite{vos94}. Here, we briefly review the most important
structural features: i) the broad density peak at contact actually consists
of two oxygen layers with hydrogen atoms in between. In this double layer
water molecules are connected by hydrogen bonds strongly tilted with respect
to the surface normal. This yields a very small distance between the oxygen
layers. ii) The detailed structure of another water double layer around
$4 \AA$, where the negative charges dominates, is not clearly resolved in the
charge density profiles. iii) The low density region between the double
layers is characterized by the pronounced double maximum of positive charge
density. Hydrogen atoms are arranged in two layers, forming hydrogen bonds
between successive water double layers. iv) The calculated water structure at
the electrode starts with a layer of oxygen atoms. Since our model electrode
acts purely repulsive on all particles, it is energetically more favourable
for water molecules immediately at the electrode to place their hydrogen
atoms into hydrogen bonds within the first water double layer. Therefore the
orientation of the first water layer is fixed by the water structure and this
yields a spontaneous net polarization or surface potential
of the electrode-water interface, even at the uncharged electrode. Due to the
strong inclination of hydrogen bonds within the water double layer, the
dipole moment of water molecules at contact is only slightly orientated
towards the bulk fluid.

The structural and electrostatic properties of the ice-like water
significantly affects the distribution of the solute in the
electrode-electrolyte interface (cf. Fig.\ref{fig5}a). It is of particular
importance for the interfacial ionic distribution how the
differently sized ions fit into the ice-like water structure. For instance,
the smaller $Na^{+}$-ion ($\sigma_{Na^{+}}=2.4 \, \AA$) penetrates into the
second water double layer around 4 \AA. The hexagonal water coordination
within a double layer is wide and flexible enough, so that the ion is able
to slip into the region between the first and second water double layer. In
this region the water structure has still enough flexibility to match the
geometry of the $Na^{+}$-solvation shell without disrupting the hydrogen
bonded network. The seperation between the first and second water double
layer increases slightly. Simultaneously positively charged hydrogens are
transfered from this interlayer region into both neighbouring water double
layers (cf Fig.\ref{fig5}b). But the
ice-like sequence of positive and negative charge density still persists. The
redistribution of positive charges can be interpreted as a reorientation of
water molecules to optimize their interaction with the neighbouring sodium
ion. Water molecules of the upper and lower double layer locally provide an
environment for the $Na^{+}$-ion, which resembles the bulk solvation shell
and is also compatible with the ice structure. Therefore the local density
$\rho_{Na}(z)$ is close to its bulk value. In the region between the first
and the second water double layer the mean electrostatic potential
(cf. Fig.\ref{fig7}) becomes repulsive for the $Na^{+}$-ion due to the large
number of positively charged hydrogen atoms arranged in hydrogen bonds
connecting successive water double layers. Although the $Na^{+}$-ion,
when further approaching the electrode, is attracted by the net dipol moment
of the first water double layer, penetration into the densely packed first
water double layer can be only accomplished by a strong, local distortion of
the ice-like ordering. But in this region the ice structure is rather rigid
and the approach of the ion is therefore strongly hindered.

The behaviour of the larger $Cl^{-}$-ion ($\sigma_{Cl^{-}}=3.24 \, \AA$) in
the electrode-electrolyte interface is quite different. The chloride ion is
too large for the hexagonal rings of hydrogen bonded water molecules,
already in the partially disordered second water double layer. In order to
provide a proper solvation environment for the $Cl^{-}$-ion hydrogen bonds
have to be disrupted, which is energetically unfavourable and therefore the
density $\rho_{Cl}(z)$ decreases in the interfacial region. Although the mean
electrostatic potential (cf. Fig.\ref{fig7}) becomes attractive for the
$Cl^{-}$-ions in the interlayer region, the rigidity of the ice-like water
structure and thus the steric repulsion increases. These competitive
interactions yields the local maximum of $\rho_{Cl}(z)$ around 4 {\AA}.
Additional chemical ion-surface or image interactions with a metal electrode
can overcome the repulsion of the rigid water layer \cite{spo93}. At a more
realistic electrode contact adsorption is expected for the larger $Cl^{-}$-ion
\cite{glo92}.

The mean electrostatic potential across the interface
\begin{equation}
\label{meanpot}
\Psi(z_{1}) = \int_{z_{1}}^{\infty} d\/z_{2} \, E\/(z_{2}) \, = \,
4 \pi \, \int_{z_{1}}^{\infty} d\/z_{2} \, (z_{1}-z_{2}) \, q\/(z_{2})
\end{equation}
is plotted in Fig.\ref{fig7} for the electrolyte and for pure water. The
shape and the magnitude of both curves are very similar throughout the entire
interface. Adding ions to the water reduces the potential drop across the
first water double layer, indicating a decreased surface dipole moment
due to the reorientation of water molecules in order to optimize their
interactions with sodium ions in the neighbourhood. The contribution of the
ionic charges to $\Psi(z)$ is rather small because of the very low
ion densities, especially near the electrode. The total dipole moment due to
the charge density in the interfacial region is given by the overall
electrostatic potential drop across the interface
\begin{equation}
\label{pzc}
\int_{\infty}^{0} d\/z \, z \, q\/(z) = \frac{\Psi(0)}{4 \pi} =
\frac{\Psi_{0}}{4 \pi},
\end{equation}
i.e. the work required to move a test charge from the bulk liquid with
$\Psi(z=+\infty)=0$ to the electrode. In the literature the potential drop
$\Psi_{0}$ at zero surface charge density is sometimes denoted as the
'potential of zero charge'. In electrochemistry the measured 'potential
of zero charge' (PZC) means a different quantity (cf. Fig.\ref{fig6}).
Often the 'surface potential' $\chi$ is used, which is related to
$\Psi_{0}$ for the uncharged electrode by the sign convention
$\chi = - \Psi_{0}$. Our value $\Psi_{0}=-74 \, mV$ for the
electrode-electrolyte interface is somewhat smaller in magnitude than
$\Psi_{0}= -110 \, mV$ calculated previously for pure water \cite{vos94}.
The negative sign of $\Psi_{0}$ reflects the preference of
the water oxygens close to the surface. Our surface potential for water
at a planar, inert wall agrees in sign with the surface potential for the
liquid-vapor interface predicted from theoretical and experimental data
\cite{wil88}, and with the surface potential obtained by Torrie and coworkers
for a water-like solvent model with multipolar hard spheres \cite{tor93}.
Recently Booth et al. have calculated $\Psi_{0}$ according to eq.(\ref{pzc})
using eq.(\ref{hnc}) with bulk correlation functions for determining
the density profiles of hydrogen and oxygen atoms for the central force water
in contact with a quasi-hard wall \cite{boo94}. Their value
$\Psi_{0} = + \, 600 \, mV$ has the opposite sign. From the corresponding
curve for the electrostatic potential, presented in Fig.6a of reference
\cite{boo94}, together with the Poisson equation, which relates the curvature
of the mean electrostatic potential to the charge density $q(z)$, it follows
a reversion of the surface dipole moment in disagreement with our and other
results \cite{tor93}. The reason for this reverted orientation is not quite
clear, because Booth and coworkers used a model and computational treatment
completely equivalent to our procedure. Only the particle-particle
correlations taken from the bulk might be different. According to our
understanding, the driving force for the spontaneous polarization of the
electrode-water interface is the tendency of the interfacial water structure
to maximize the number of hydrogen bonds \cite{vos94}. \\*

\noindent
{\bf C. Results for the charged electrode.} In this section we discuss
the influence of an external electric field due to an electrode covered with
a constant charge density on the distribution of ions in the
electrode-electrolyte interface. The surface charge density $\omega$ is
increased up to $\omega = \pm \, 10.4 \, \mu\/C / {cm}^{2}$ corresponding to
an electric field strength of $1.18 \, \times \, 10^{8} \, V/cm$, comparable
to a very strong electric field applicable in electrochemical experiments. Our
previous calculations for pure water have shown that the ice-like water
structure near the electrode is preserved under the influence of such field
strengths. The reorientation and the exchange of hydrogen atoms between the
two positions in a hydrogen bond provides for enough polarizability. In case
of a small negative surface charge density alignment of
water dipoles with respect to the electric field is hindered. For the
hydrogen atoms of the contact water layer it is energetically more favourable
to remain still between the oxygen layers as long as the energy gain in the
external potential is smaller than the binding energy in the hydrogen bond.
Therefore the polarizability decreases. Only when the hydrogen bond energy is
overcompensated by the potential energy gain of the hydrogen atom in the
external field, some hydrogens follow the attractive electric field.
As a result the interfacial polarizability increases.
This reorientation of water molecules starts at
$\omega = - \, 3.7 \, \mu\/C / {cm}^{2}$ \cite{vos94}. Experimental and
theoretical investigations found a maximum of the interfacial entropy
for surface charge densities in the range from
$- \, 4$ to $-6 \, \mu C / {cm}^{2}$ \cite{gui86}, where the hydrogen atoms
of the contact water layer, according to our calculations, can occupy two
different positions. The mechanism of water reorientation still holds for the
water solvent in the electrode-electrolyte interface. The ion density
profiles, presented in Figure \ref{fig8}, have to be discussed with respect
to this specific behaviour of water molecules in an applied electric field.
For a surface charge density $\omega > 0$ the electric field cannot force
the large $Cl^{-}$-ion to penetrate into the contact water layer, whereas
an electric field of opposite sign but of same magnitude brings the
$Na^{+}$-ion close to the electrode surface. This different ability of
approaching the electrode is a result of two competing forces: the
attraction due to the external electric field and the steric repulsion of
the differently sized ions by the ice-like water matrix established in front
of the electrode. When the external field is attractive enough for the
hydrogen atoms of contact water molecules to flip towards the electrode, the
$Na^{+}$-ions finally invade the first water double layer and even get into
contact with the surface (cf. dashed-dotted line in Fig.\ref{fig8}a).
A measure for the excess of ions in the electrode-electrolyte interface is
\begin{equation}
\Gamma_{\alpha} = \int_{0}^{\infty} d\/z \, [ \, \rho_{\alpha}(z) -
\rho_{\alpha} \, ].
\end{equation}
The plot of $\Gamma_{\alpha}$ for the $Na^{+}$-ions vs $- \omega$ in the
inset of Fig.\ref{fig8}a reflects the rapidly increasing penetration of
$Na^{+}$-ions into the surface water structure beyond
$\omega \approx -4 \, \mu\/C / {cm}^{2}$, where the pronounced water
reorientation becomes significant. Contact adsorption of $Na^{+}$-ions has
been also observed in computer simulations \cite{ros91}.

In order to elucidate the effect of the asymmetric response of ionic charges
in an applied electric field on the differential capacitance
$C = \partial\/\omega / \partial\/\Psi_{0}(\omega)$, we evaluated
$\Psi_{0}(\omega)$ across the interface for a series of surface charge
densities. Some of these values $\Psi_{0}$ are summarized in table \ref{tab5}.
For comparision the last column of table \ref{tab5} shows the values of
$\Psi_{0}(\omega)$ for pure water. As already mentioned, at zero surface
charge the electrostatic potential drop across the interface is only slightly
smaller than for pure water, but the negative sign is retained. The same
trend was observed by Torrie and coworkers  when adding ions to their
multipolar, water-like solvent model \cite{tor93}. In the third column we
have listed the change $\Delta\/\Psi_{0}=\Psi_{0}(\omega)-\Psi_{0}(\omega=0)$.
Only for small surface charge densities $|\Delta\/\Psi_{0}|$ behaves like a
symmetric function of $\omega$, whereas for large negative surfaces charges
$|\Delta\/\Psi_{0}|$ grows more strongly than for corresponding $\omega > 0$.
This implies a smaller differential capacitance for the negatively charged
electrode. In order to decide whether this asymmetry is due to the response
of the water solvent or due to the ions, we have examined the water and ionic
contributions to the mean electrostatic potential $\Psi_{0}(\omega)$
seperately, as shown in table \ref{tab5}. For the 0.01 molar solution the
ionic contribution $\Psi_{ionic}$ is rather small in comparision with
$\Psi_{water}$. For almost all surface charge densities $\Psi_{ionic}$ is
independent of $\omega$. Only the pronounced adsorption of $Na^{+}$-ions
yields a distinct increase in $\Psi_{ionic}$. Since $\Psi_{ionic}$ is rather
small and flat, the asymmetric response of the total $\Psi_{0}(\omega)$ is
primarily a property of the interfacial water structure. Nethertheless, to
some extent $\Psi_{water}$ is indirectly affected by the presence of ions
through a restructuring of the contact double water layer, as emphasized by
comparing the solvent component $\Psi_{water}$ and $\Psi_{0}(0 \, M)$ for
pure water. $\Psi_{water}$ for the electrolyte varies less strongly with
respect to $\omega$. As a result the differential capacitance for the
electrode-electrolyte interface is larger than for pure water over the whole
range of considered surface charge densities.

The differential capacitance
$C=\partial\/\omega / \partial\/\Psi_{0}(\omega)$ obtained from the
derivative of a cubic spline fit to the $\Psi_{0}(\omega)$ data is shown
in Fig.\ref{fig9} as a function of surface charge density $\omega$. The
curves for the 0.01 molar aqueous solution of sodium chloride and for pure
water, taken from \cite{vos94}, are very similar. Because
$\partial\/\Psi_{ionic}(\omega) / \partial\/\omega$ is negligible in
comparision with $\partial\/\Psi_{water}(\omega) / \partial\/\omega$,
the total capacitance of the electrode-electrolyte interface is dominated by
the response of the interfacial water structure. The minimum in the
differential capacitance curve has been identified with the beginning of
water reorientation in the adsorbed water layer \cite{vos94}. The asymmetry
in the differential capacitance curve with respect to zero surface charge and
the existence of a minimum at negative surface charge density are typical of
the experimental data for simple electrolytes \cite{gra52}. It has been
argued that the asymmetry is a consequence of ionic polarizability or of
ion-specific interaction with the electrode\cite{wei93}. Our result suggest
that the minimum can be explained by the asymmetric water response near an
electrode, at least for low concentrated electrolytes. We found the
differential capacitance for the electrolyte larger than for pure water over
the whole range of surface charges considered, indicating an overall larger
polarizability of the interfacial structure. From the rather small direct
effect of the ion distribution on the differential capacitance, we infer
that due to the presence of sodium and cloride ions in the interface water
molecules are already reorientated and less strongly fixed within the
ice-like water structure yielding a larger polarizability of the interfacial
water structure.

\section{Conclusions}
\noindent
We have extended our previously developed integral equation method for
calculating bulk correlation functions and density profiles for the
central force water model to an detailed microscopic model of an aqueous
electrolyte. Spherical symmetric pair potentials for the additional ion-ion
and ion-water interactions are designed to represent the different size of
sodium and chloride ions as well as the orientation of water molecules around
both ions. Only for the water correlation functions we have added to the
HNC-approximation appropriate bridgefunctions, which are optimized in order
to yield the bulk water structure \cite{vos94}. The calculated ion-water
pair distribution functions reflects specific features of the ionic hydration
shells in aqueous solutions. On average six water molecules are located in the
first $Na^{+}$-hydration shell and their dipole moment is tilted with respect
to the ion-oxygen radial vector. The water molecules around the larger
$Cl^{-}$-ion are bound more weakly and therefore the first hydration shell of
the anion is less sharply defined. We have found that in the anionic
hydration shell the water molecules are orientated with one of the hydrogen
atoms towards the $Cl^{-}$-ion.

Taking these correlation functions of the homogeneous fluid for describing the
particle-particle correlations near the electrode, we have
calculated the density profiles for the various particle species near a
planar electrode, which is neutral or charged and has some hard wall
repulsion, equal for all particles. The charge density profile of the water
solvent that is established at a neutral electrode resembles that for pure
water, which has been interpreted in terms of an ice-like arrangement of
water molecules \cite{vos94}. This layered structure extends about two water
double layers into the liquid. Water molecules adjacent to the surface are
slightly orientated toward the bulk fluid and thus the interfacial water
structure is polarized, even without any external electric field. This
specific water structure induces an asymmetric ion distribution in front
of the electrode depending strongly on the ion size. The smaller $Na^{+}$-ion
can easily penetrate into the second water double layer and even comes into
contact with the surface for sufficiently attractive fields.
Penetration of the larger $Cl^{-}$-ion into this interfacial region is
strongly hindered. Strong local, steric interactions between the ions and
the ice-like water structure dictates this different behaviour. Electrostatic
effects are only of minor importance. The cavities formed by the puckered
hexagonal water rings within a water double layer are wide and flexible
enough to match the size of the smaller $Na^{+}$-ion. Water molecules are
reorientated without disrupting the ice structure in order to provide an
solvation environment for the $Na^{+}$-ion. This reorientation of water
molecules is responseable for a reduction of the surface potential from
$\Psi_{0} = - \, 110 \, mV$ in case of pure water to
$\Psi_{0} = - \, 74 \, mV$ calculated for the 0.01 molar solution of sodium
chloride at the uncharged electrode. The $Cl^{-}$-ion is too large for the
rigid hexagonal water cavities and therefore the $Cl^{-}$-ions are strongly
repelled.

Under the influence of an external electric field of reasonable strength
the ice-like water structure near the electrode is still preserved. Chloride
ions are less effectively attracted by a positively charged electrode than
sodium ions by an equivalently attractive electric field because the rigid
first water double layer resists the electric field.  On the other hand
$Na^{+}$-ions suddenly pentrate into the contact water layer, when the
electrode charge is large enough ($- 4 \, \mu C /{cm}^{2}$) to
induce 'flipping' of adsorbed water molecules within the ice structure.

Although sodium and chloride ions behaves differently at oppositely charged
electrodes, this has a negligible effect on the interfacial differential
capacitance $C = \partial\/\omega / \partial\/\Psi_{0}(\omega)$. At least
for dilute aqueous electrolytes the asymmetry in the differential
capacitance curve with respect to zero surface charge density is solely
determined by the ice-like water structure as discussed more extensively in
reference \cite{vos94}. Due to the presence of $Na^{+}$-ions some water
molecules are less strongly fixed within the interfacial water structure and
already reorientated in order to optimize their interaction with the
neighbouring ion. As a result, we have obtained a larger polarizability and
therefore a larger differential capacitance for the electrode-electrolyte
interface compared with pure water. Our present model has to be further
developed for instance by introducing specific electrode-ion interactions in
order to understand more features of the experimental differential capacitance
curves microscopically.

\acknowledgements
\noindent
Helpful discussions with M. Kasch and J.E. Diaz-Herrera and financial support
of this work by Deutsche Forschungsgemeinschaft are gratefully acknowledged.

\begin{figure}
\raggedright
\caption{a) Pair distribution functions $g_{NaH}(r)$ (-----) and
$g_{NaO}(r)$ (-- --) for a 0.01 molar aqueous NaCl-solution.
b) The corresponding coordination numbers $N_{NaH}(r)$ (-----) and
$N_{NaO}(r)$ (-- --).}
\label{fig1}
\end{figure}
\begin{figure}
\raggedright
\caption{a) Pair distribution functions $g_{ClH}(r)$ (-----) and
$g_{ClO}(r)$ (-- --) for a 0.01 molar aqueous NaCl-solution.
b) The corresponding coordination numbers $N_{ClH}(r)$ (-----) and
$N_{ClO}(r)$ (-- --).}
\label{fig2}
\end{figure}
\begin{figure}
\raggedright
\caption{a) The potential of mean force for the ion pair $Na^{+}-Cl^{-}$ and
b) the bare interaction potential $u_{NaCl}(r)$.}
\label{fig3}
\end{figure}
\begin{figure}
\raggedright
\caption{The potential of mean force for the ion pairs $Na^{+}-Na^{+}$
(-----) and $Cl^{-}-Cl^{-}$ (-- --).}
\label{fig4}
\end{figure}
\begin{figure}
\raggedright
\caption{a) The density profile $\rho_{\alpha}(z)/\rho_{\alpha}$ for hydrogen
(-----) and oxygen atoms (-- --), for $Na^{+}$- ($\cdots$) and
$Cl^{-}$-ions (-- $\cdot$ --) at the uncharged electrode. b) The charge
density $q(z)=q[\rho_{H}(z)-2\rho_{O}(z)]$ for the water solvent in the
electrolyte (-----) and for pure water (-- --).}
\label{fig5}
\end{figure}
\begin{figure}
\raggedright
\caption{The mean electrostatic potential $\Psi(z)$ near an
uncharged electrode for pure water \mbox{(-- $\cdot$ --)}, the 0.01
molar aqueous solution of sodium chloride \mbox{(-----)}, ionic
\mbox{(-- --)} and solvent contributions \mbox{($\cdots$)}.}
\label{fig7}
\end{figure}
\begin{figure}
\raggedright
\caption{Schematic diagram of contributions to the electrochemical potential
of an electron. Different contributions are plotted as \mbox{$-e \phi\/(z)$}.
Electrostatic potential \mbox{({\bf -----})}, potential in front of liquid
surface \mbox{({\bf -- --})}, \mbox{$\Psi_{0}$}: the surface potential of
the electrolyte eq.(\protect{\ref{pzc}}), PZC: potential of zero charge
measured in electrochemistry, \mbox{$\Phi_{i}$}: workfunction of the
electrodes, \mbox{$\chi_{M}$}: surface potential of metal, S: total surface
potential at reference electrode, AP: absolute electrode potential at
reference electrode related to redox reaction.}
\label{fig6}
\end{figure}
\begin{figure}
\raggedright
\caption{The density profiles for the $Na^{+}$- and $Cl^{-}$-ions at the a)
negatively and b) positively charged electrode for surface charge densities
$\omega = 0$ (-----), $\pm 1.6$ (-- --), $\pm 3.2$ ($\cdots$) and $\pm 6.4 \,
\mu C / {cm}^{2}$ (-- $\cdot$ --). The inset shows $\Gamma_{Na}(\omega)$ as
a function of $- \omega$.}
\label{fig8}
\end{figure}
\begin{figure}
\raggedright
\caption{The differential capacitance for the 0.01 molar aqueous NaCl-solution
and for pure water.}
\label{fig9}
\end{figure}
\newpage
\renewcommand{\thefootnote}{\alph{footnote}}
\setcounter{table}{1}
\setcounter{footnote}{0}

\begin{table}
\caption{\label{tab2}Structural properties of the ionic hydration shell at
T=300 K.}
\centerline{
\squeezetable
\begin{minipage}{13cm}
\begin{tabular}{|l|d|d|}
          & $Na^{+}$ & $Cl^{-}$ \\ \hline
          & \multicolumn{2}{l|}{ion-oxygen distance in {\AA}} \\
This work    & 2.34 & 3.52 \\
Smith et al.\footnote{MD, infinite dilution} \cite{smi92} & 2.33 & 3.40 \\
Pettitt et al.\footnote{RISM, infinite dilution} \cite{pet86} & 2.30 & 3.45 \\
X-ray \cite{new89} & 2.38 - 2.40 & - \\
neutron diffraction \cite{nei85} & - & 3.2 - 3.34 \\ \hline
          & \multicolumn{2}{l|}{ion-hydrogen distance in {\AA}} \\
This work & 2.69 and 3.21 & 2.78 \\
Smith et al. \cite{smi92}   & 3.0 & 2.53 \\
Pettitt et al. \cite{pet86}  & 3.05 & 2.0 \\
neutron diffraction \cite{nei85} & - & 2.2 - 2.26 \\ \hline
          & \multicolumn{2}{l|}{coordination number ion-oxygen} \\
This work & 6.5 & 17.9 \\
Bopp et al.\footnote{MD, 2.2 m NaCl} \cite{bop79} & 5.9 & 8.4 \\
Thuraisingham et al.\footnote{HNC, 0.069M NaCl} \cite{thu83} & 15.9 & 17.3 \\
Ichiye et al.\footnote{RHNC, infinite dilution} \cite{ich88} & 5.7 & - \\
Pettitt et al. \cite{pet86} & 4.28 & 11.73 \\
X-ray \cite{new89}  & 4.6 & - \\ \hline
          & \multicolumn{2}{l|}{coordination number ion-hydrogen} \\
This work & 14.9 & 18.2 \\
Ichiye et al. \cite{ich88}  & -    & 14.8 \\
Pettitt et al. \cite{pet86}   & 6.97 &  -   \\
neutron diffraction \cite{nei85} & - & 5.3 - 6.2
\end{tabular}
\end{minipage}
}
\end{table}
\newpage
\setcounter{table}{0}
\begin{table}
\caption{\label{tab1}Ion-water Lennard-Jones parameters. $\sigma$ in {\AA}
and $\epsilon$ in $10^{-12}erg$.}
\centerline{
\begin{minipage}{8cm}
\begin{tabular}{|c|d|d|d|}
$\alpha$ & $\sigma_{\alpha\/O}$
         & $\sigma_{\alpha\/H}$
         & $\epsilon_{\alpha\/O}=\epsilon_{\alpha\/H}$ \\ \hline
$Na^{+}$ & 2.222  & 1.547 & 0.1853 \\
$Cl^{-}$ & 3.378  & 2.703 & 0.03708 \\
\end{tabular}
\end{minipage}
}
\vspace{1cm}
\setcounter{table}{2}
\caption{\label{tab3}Orientation of water molecules, mean values of $\Theta$
in degree.}
\centerline{
\begin{minipage}{8cm}
\begin{tabular}{|c|c|c|}
         & $Na^{+}$ & $Cl^{-}$ \\ \hline
This work & 117 & 44 \\
Smith et al.\footnote{MD with CF-water, from $g_{\alpha\beta}(r)$}
\cite{smi92} & 124 & 34 \\
Heinzinger et al.\footnote{MD with CF-water,
from $\Theta={\cos}^{-1}\/\langle \cos\/\Theta \rangle$} \cite{bop79}
& 139 & 55 \\
Heinzinger et al.\footnote{same as b with the ST2-water}
\cite{bop79} & 132 & 54 \\
\end{tabular}
\end{minipage}
}
\vspace{1cm}
\caption{\label{tab4}Energy of solvation in $eV$.}
\centerline{
\begin{minipage}{10cm}
\begin{tabular}{|d|d|d|}
                     & $U_{Na^{+}}^{solv}$ & $U_{Cl^{-}}^{solv}$ \\ \hline
This work            & -8.60        & -7.34       \\
Chandrasekhar\footnote{MD, 0.444 molal} et al. \cite{cha84} & -8.47 $\pm$ 0.04
                                               & -6.21 $\pm$ 0.04 \\
Zhu\footnote{MD, 1.791 molal} et al. \cite{zhu92} & -6.92 $\pm$ 0.53
                                     & -4.51 $\pm$ 0.53 \\
Pettitt\footnote{RISM, infinite dilution} et al. \cite{pet86} & -7.21 & -8.25
\\
\end{tabular}
\end{minipage}
}
\newpage
\caption{\label{tab5}Surface potential $\Psi_{0}$,
$\Delta\/\Psi_{0} = \Psi_{0}(\omega) - \Psi_{0}(\omega=0)$, ionic and
solvent contributions in volts for the 0.01 M aqueous NaCl-solution and
$\Psi_{0}(0 \, M)$ for pure water.}
\centerline{
\begin{minipage}{13cm}
\begin{tabular}{|d|d|d|d|d|d|}
$\omega \, [\mu C / {cm}^{2}]$ & $\Psi_{0}$ &
$\Delta\/\Psi_{0}$ & $\Psi_{water}$ & $\Psi_{ionic}$ & $\Psi_{0}(0 \, M)$
\\ \hline
-10.4 & -1.039 & -0.9652 & -1.018 & -0.0210 & -1.0890 \\
-6.4  & -0.670 & -0.5960 & -0.657 & -0.0130 & -0.7140 \\
-3.2  & -0.371 & -0.2973 & -0.359 & -0.0117 & -0.4119 \\
-2.4  & -0.296 & -0.2227 & -0.285 & -0.0116 & -0.3364 \\
-1.6  & -0.222 & -0.1483 & -0.210 & -0.0115 & -0.2608 \\
0.0   & -0.074 &  0.0000 & -0.062 & -0.0114 & -0.1100 \\
1.6   &  0.073 &  0.1470 &  0.085 & -0.0114 &  0.0402 \\
2.4   &  0.146 &  0.2199 &  0.158 & -0.0114 &  0.1149 \\
3.2   &  0.219 &  0.2924 &  0.230 & -0.0114 &  0.1895 \\
6.4   &  0.504 &  0.5775 &  0.516 & -0.0114 &  0.4847 \\
10.4  &  0.849 &  0.9226 &  0.860 & -0.0112 &  0.8464 \\
\end{tabular}
\end{minipage}
}
\end{table}

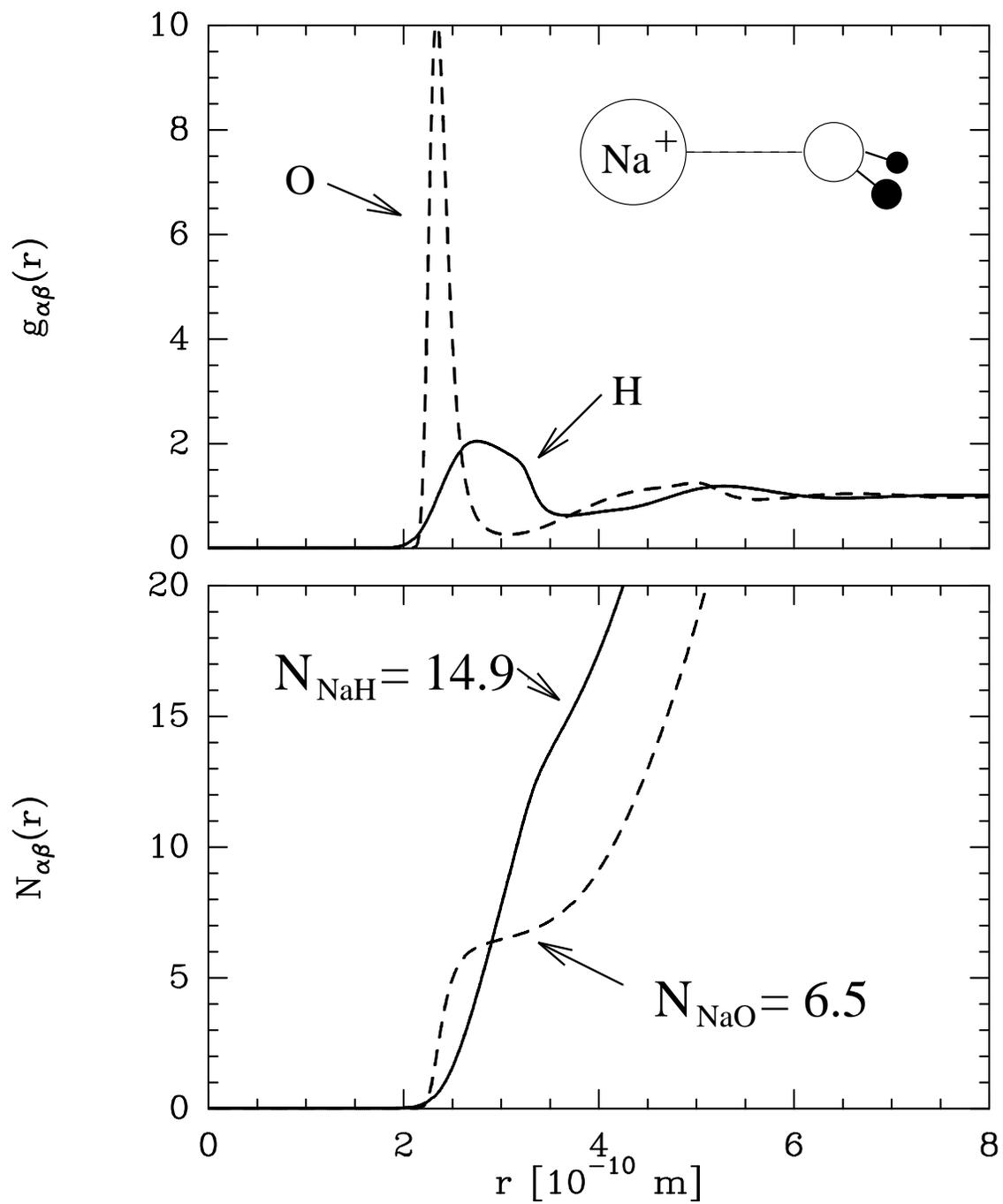
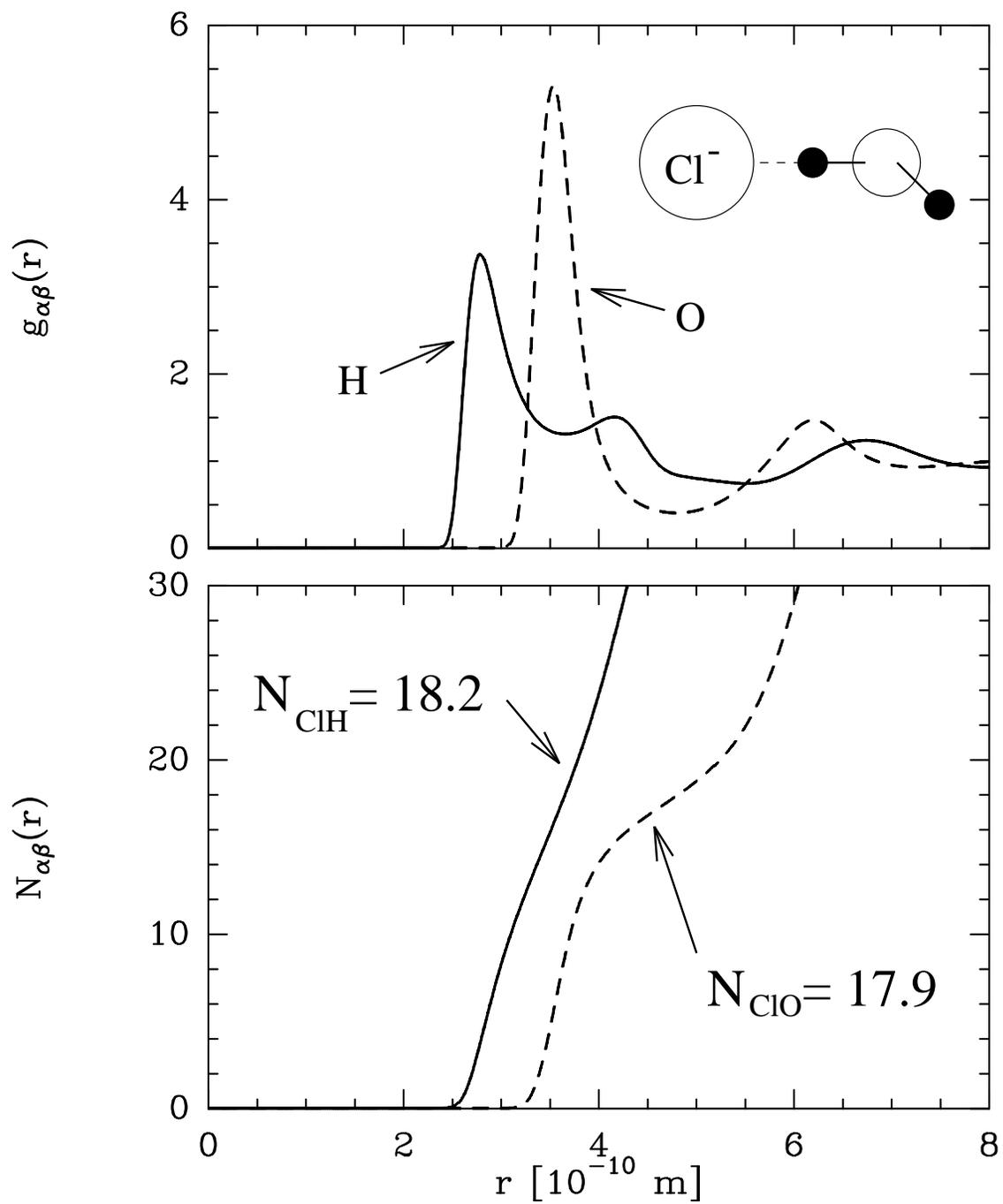
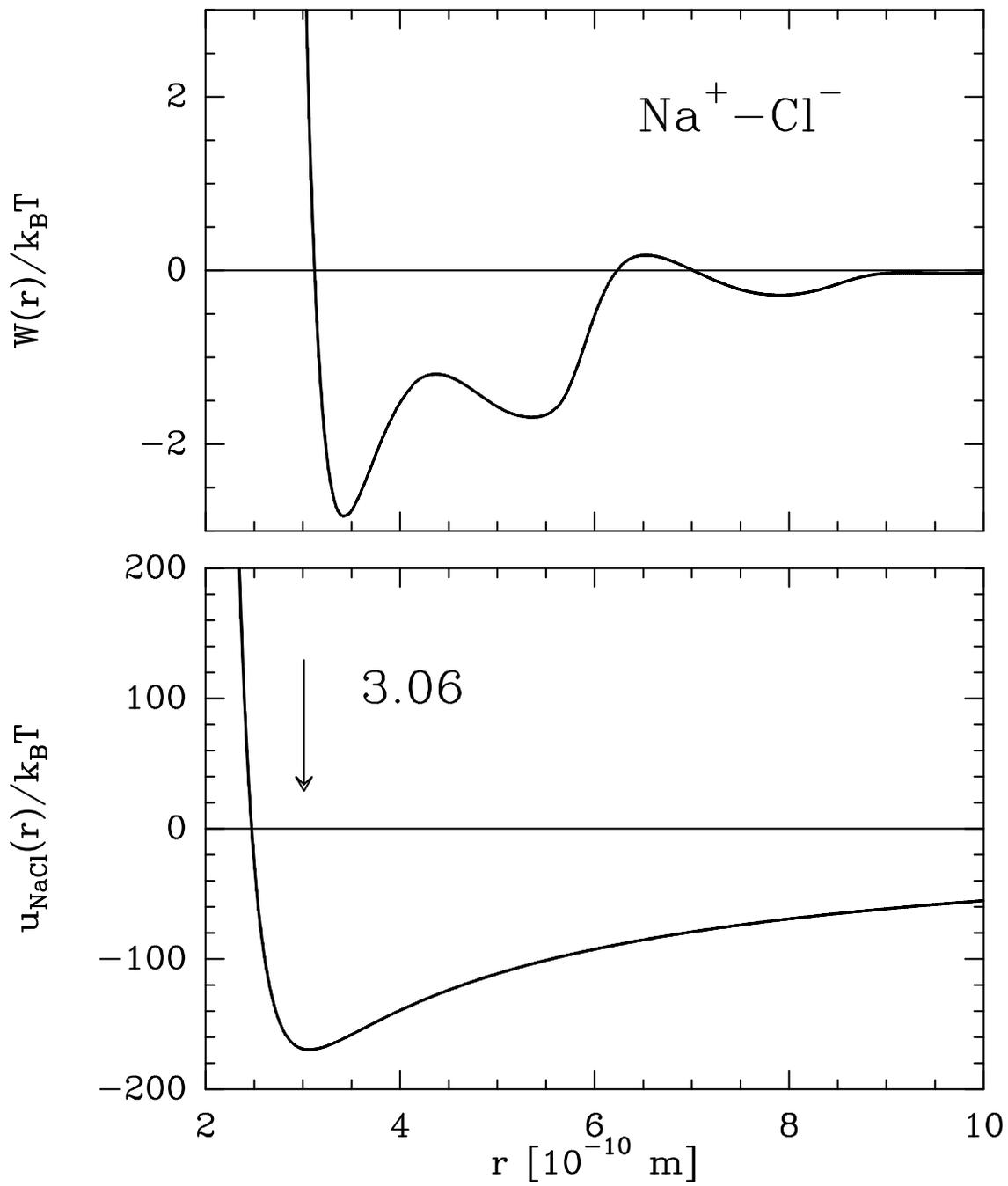
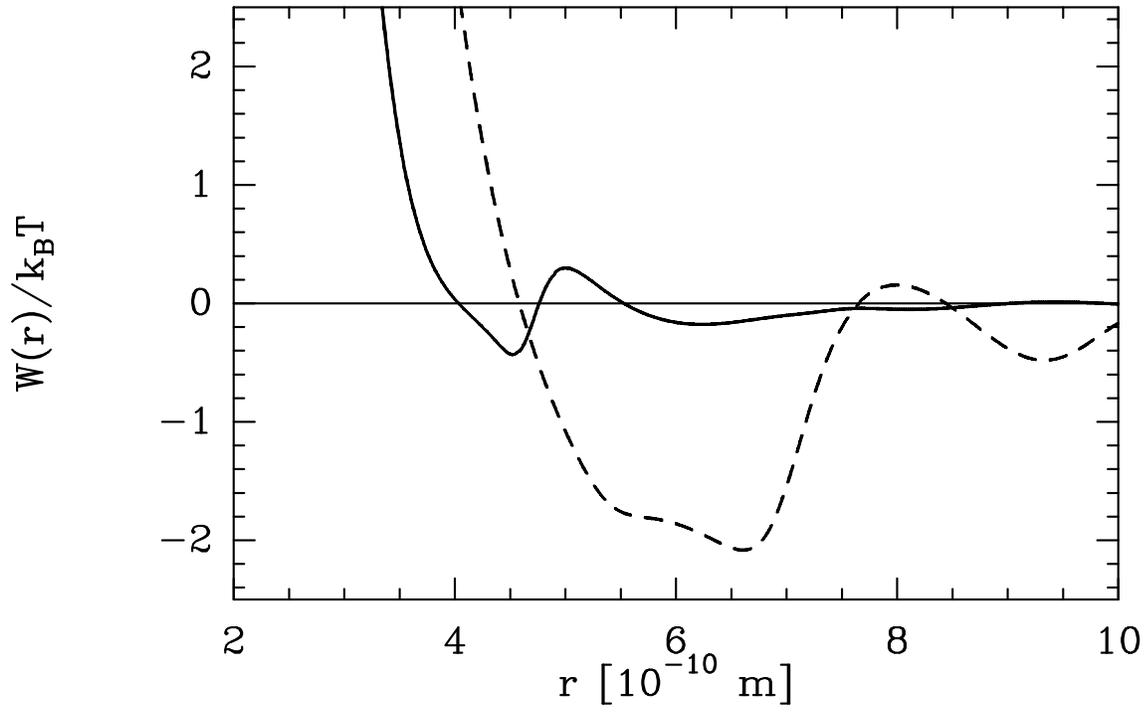
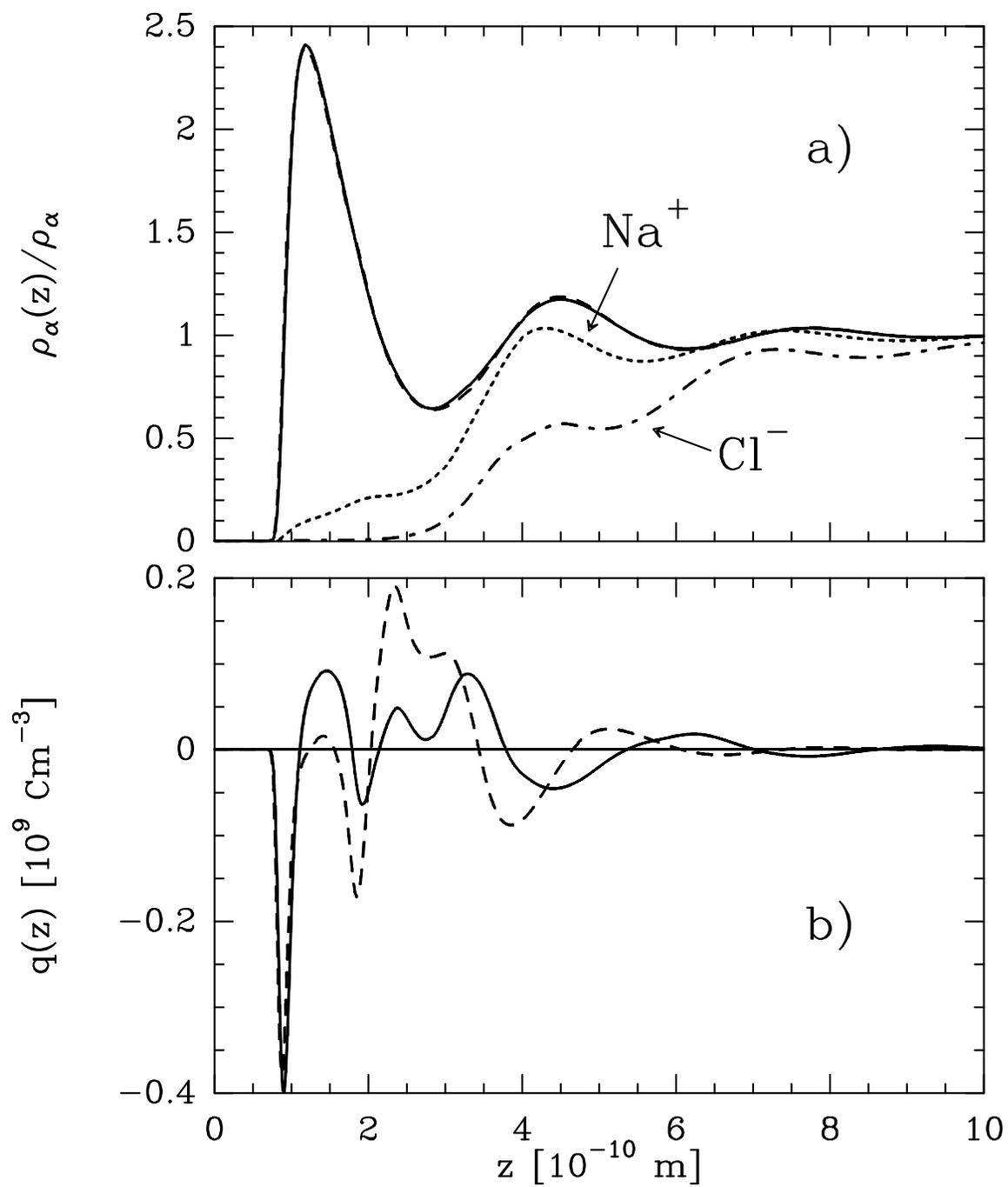
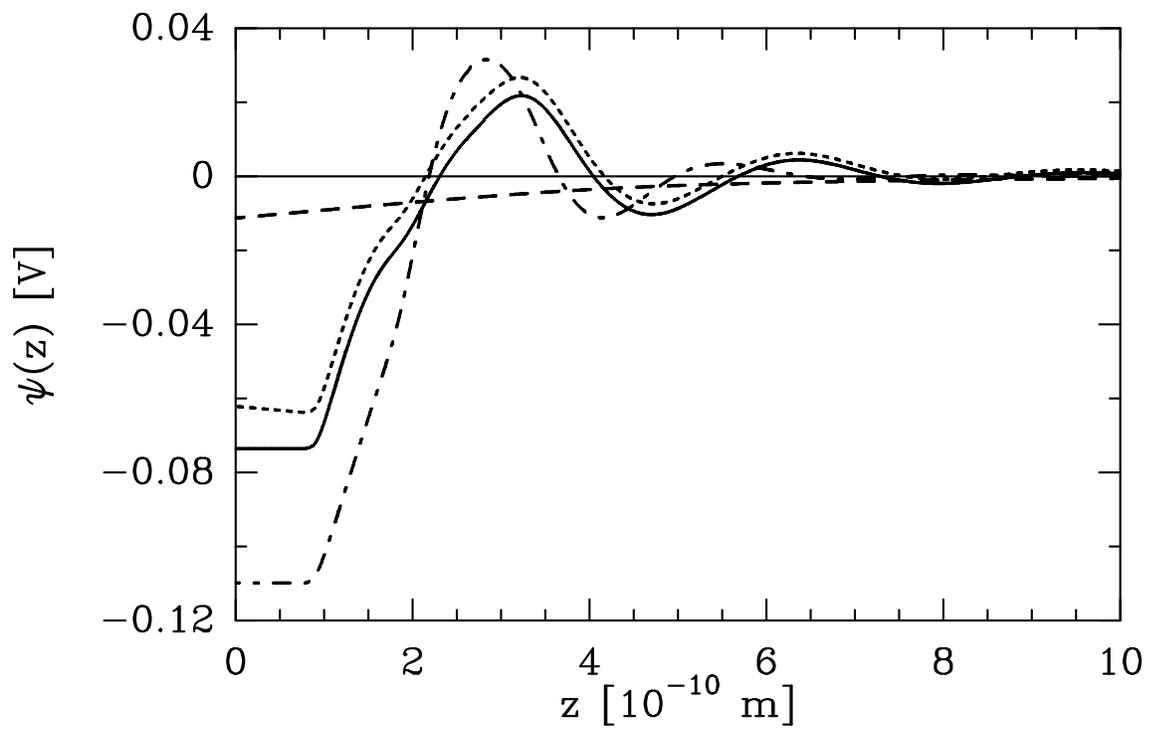
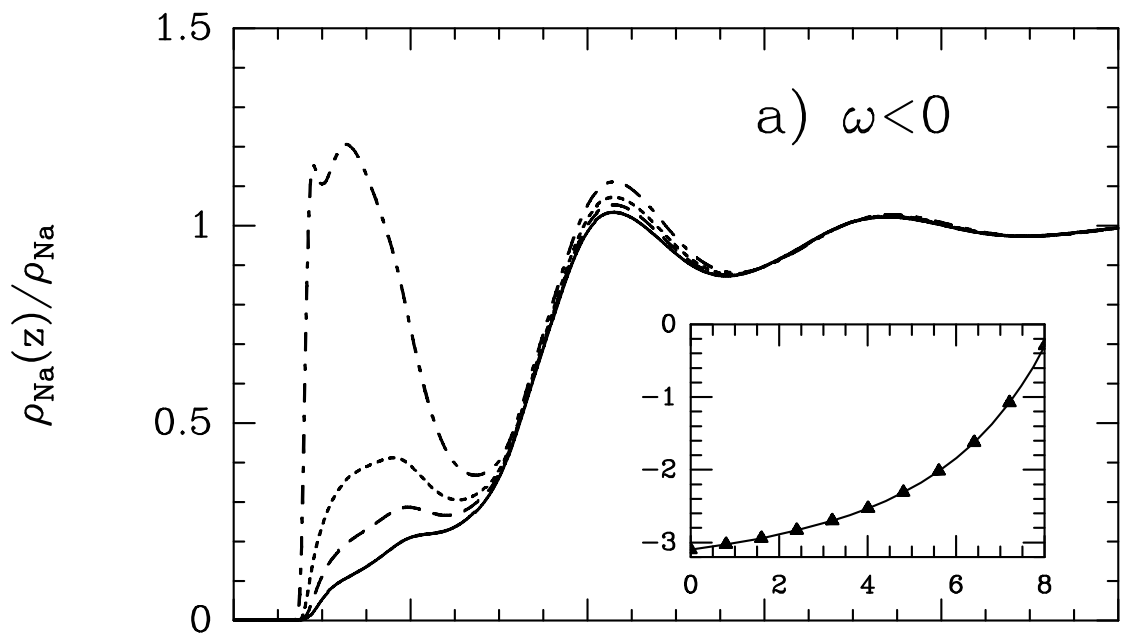
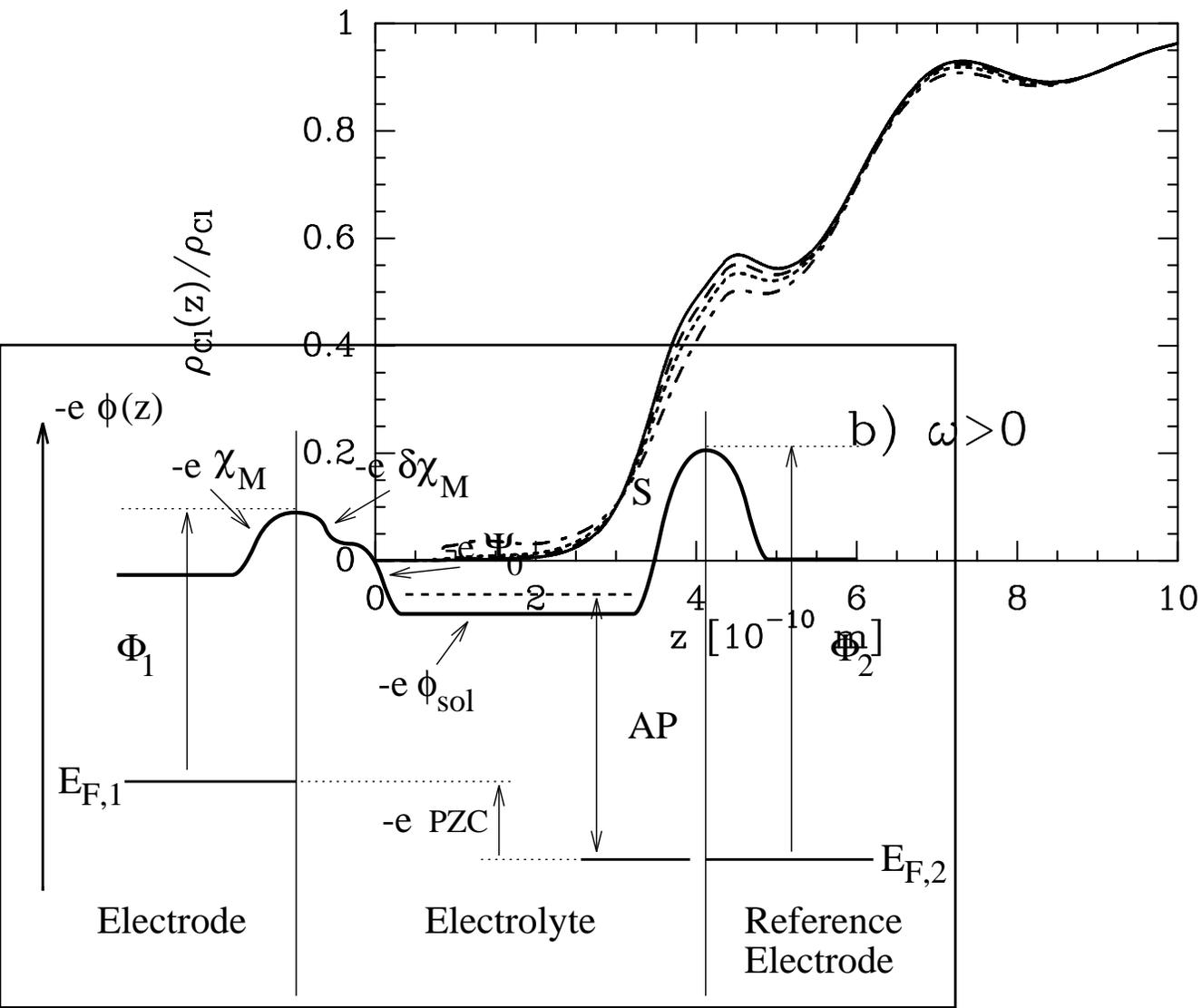
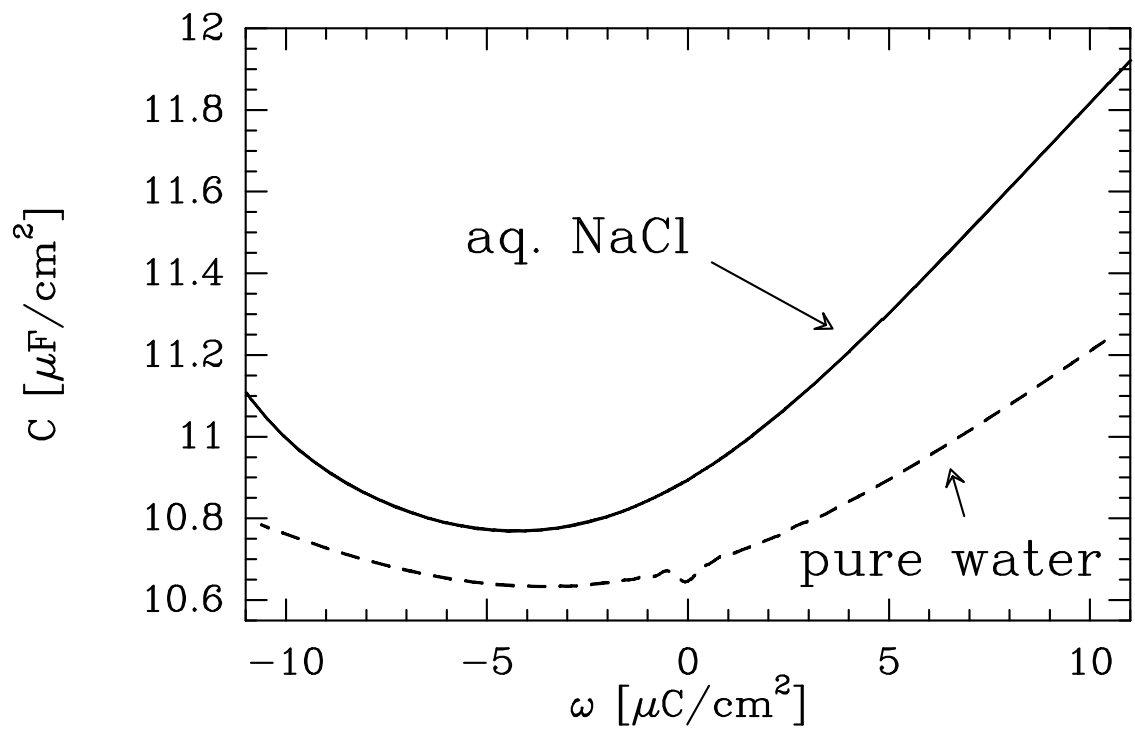
\end{document}